\documentclass[prb,aps,showpacs,twocolumn,superscriptaddress,10pt,amsmath,amssymb]{revtex4-1}
\usepackage{epsfig}
\usepackage{amsmath}
\usepackage{amssymb}
\usepackage{graphicx}
\usepackage{enumerate}
\usepackage{epstopdf}
\usepackage{circuitikz}
\usepackage{units}

\newcommand*{\bb}[1]{\boldsymbol{#1}}

\makeatletter 

\begin{document}

\title{Thermal conductivity changes across a structural 
       phase transition: the case of high-pressure silica}

\author{Hugo Aramberri}
\affiliation{Institut de Ci\`{e}ncia de Materials
             de Barcelona (ICMAB--CSIC), Campus de Bellaterra,
             08193 Bellaterra, Barcelona, Spain}

\author{Riccardo Rurali}
\affiliation{Institut de Ci\`{e}ncia de Materials
             de Barcelona (ICMAB--CSIC), Campus de Bellaterra,
             08193 Bellaterra, Barcelona, Spain}

\author{Jorge \'I\~niguez}
\affiliation{Materials Research and Technology Department, 
             Luxembourg Institute of Science and Technology, 
             5 avenue des Hauts-Fourneaux,
             L-4362 Esch/Alzette, Luxembourg}

\date{\today}
\begin{abstract}
By means of first-principles calculations, we investigate the thermal properties of 
silica as it evolves, under hydrostatic compression, from a stishovite phase into a 
CaCl$_2$-type structure. We compute the thermal conductivity tensor by solving the 
linearized Boltzmann transport equation iteratively in a wide temperature range, 
using for this the pressure-dependent harmonic and anharmonic interatomic couplings 
obtained from first principles. Most remarkably, we find that, at low temperatures, 
SiO$_2$ displays a large peak in the in-plane thermal conductivity and a highly 
anisotropic behavior close to the structural transformation. We trace back the origin 
of these features by analyzing the phonon contributions to the conductivity.
 We discuss the implications 
of our results in the general context of continuous structural transformations in 
solids, as well as the potential geological interest of our results for silica.
\end{abstract}
\maketitle

\section{Introduction}
\label{sec:intro}

The thermal properties of solids and nanostructures are attracting 
a growing interest, both from the fundamental standpoint and 
for application-driven research. The quest for low thermal 
conductivity materials, for instance, is one of the main challenges 
in the development of efficient thermoelectric devices~\cite{thermoel},
while materials with a large thermal conductivity could help overcome 
the problem of heat dissipation at the nanoscale, which has become 
one of the major hitches for nanoscaling electronic 
devices~\cite{chen2005nanoscale,luo2013nanoscale}.
The thermal properties of an insulating solid are governed by its 
lattice thermal conductivity. This physical property is determined 
by the lattice vibrational modes (phonons) and by the scattering 
processes they encounter. At structural phase transitions which are 
driven by soft phonons, the lattice thermal conductivity is expected 
to be strongly modified, since the soft mode will experience a frequency 
and group velocity shift which will in turn modify the allowed phonon 
scattering processes in the system.

In this article we investigate the thermal properties of SiO$_2$  
close to a structural phase transition.
Some theoretical works have already calculated the thermal properties
from first-principles for a handful of oxides such as ferroelectric
PbTiO$_3$~\cite{ARoy}, MgO~\cite{MgO1,MgO2,dekura2017}, and MgSiO$_3$~\cite{dekura2013,MgSiO3},
but none of them was studied close to a structural phase transition.

Silicon dioxide, also known 
as silica, is not only the most abundant compound on Earth~\cite{morgan1980chemical}, 
the Moon~\cite{Kuskov1997239} and the terrestrial planets~\cite{morgan1980chemical,
rieder1997chemical}, but is also at the heart of modern day electronic devices
and is widely used as a substrate in thin film growth. The phase diagram 
of SiO$_2$  is very rich, with a wide variety of silica polymorphs such as 
$\alpha$-quartz, $\beta$-quartz, cristobalite, tridymite or coesite. 
In these phases Si shows a tetrahedral coordination with the surrounding 
O atoms. At high pressures (of the order of gigapascals) the coordination 
of silicon in SiO$_2$  becomes octahedral, giving rise to the stishovite 
phase, the CaCl$_2$-type phase, or the $\alpha$-PbO$_2$-type phase at even 
higher pressures (of the order of 100~GPa). In particular, the pressure 
induced phase transition from the stishovite to the CaCl$_2$-type phase is
considered as a paradigmatic pseudoproper ferroelastic phase transition.
While the atomic structure~\cite{Spackman1987,PhysRevLett.80.2145},
phonon band structure~\cite{togo2008first,tsuchida1989new,cohen1993first,Oganov,tsuchiya}, 
electronic structure~\cite{Alvarez199837,li1994x,Soldatov2000687}, 
and thermodynamic potentials describing the structural phase 
transition~\cite{Carpenter,carpenter2006,Andrault720,togo2008first} 
have been widely studied, both experimentally and theoretically,
no previous work has addressed the thermal properties of the stishovite 
nor the CaCl$_2$-type phases. In this work we calculate the thermal 
properties of the stishovite and CaCl$_2$-type high pressure phases of 
silica as a function of temperature and pressure from first principles.

\section{Computational Methods}
\label{sec:model}

\subsection{Density functional calculations of the interatomic force constants}

We performed first-principles electronic structure calculations 
within density functional theory (DFT), using the Vienna Ab-initio Simulation 
Package~\cite{VASP-Kresse2-PhysRevB.48.13115,VASP-Kresse199615} 
(\textsc{VASP}) along with the Perdew-Burke-Ernzerhof~\cite{pbegga} 
(PBE) implementation of the generalized gradient approximation (GGA) for
the exchange--correlation functional.
We employed a plane wave basis set with a 500~eV kinetic energy cutoff
with the projector augmented-wave method~\cite{BlochlPRB94,KressePRB99}.
For the ground state calculations we considered the primitive cells of
the stishovite and the CaCl$_2$-type phases (see Figure~\ref{crystal}) and
carefully optimized the lattice vectors and the atomic position
until the residual stress and the forces were smaller than 10$^{-2}$ kbars and 
10$^{-6}$ eV/\AA, respectively. The Brillouin zone (BZ) was sampled
with a converged 4$\times$4$\times$6 Monkhorst-Pack~\cite{MonkhorstPRB76} 
grid of {\bf k}-points. Hydrostatic pressure was applied by varying
the lattice vectors as described below, and allowing the atomic
coordinates to relax.

We employed the direct supercell approach to obtain the phonon band 
structures. In this approach the second-order interatomic force 
constants (IFCs) are computed directly by considering supercells of 
the corresponding primitive cell with small enough (0.01~\AA) atomic 
displacements. We employed the \textsc{phonopy} 
software~\cite{togo2015first} to generate the minimal set of 
supercells required to obtain the IFCs, while \textsc{VASP} was 
used to compute the Hellmann-Feynman forces in these cells.
The long-range corrections to the potential were included to correctly 
address for dipole-dipole interactions arising from longitudinal 
optic (LO) vibrational modes of the crystal.
The supercells with atomic displacements employed to calculate the 
third order IFCs, which account for the three-phonon scattering 
processes, were generated using \textsc{thirdorder.py}~\cite{shengbte}.
We employed 3$\times$3$\times$4 supercells for computing both the 
second and third order IFCs at each pressure, since we found that the 
subsequent lattice thermal conductivity is 
 well converged when compared to the results obtained for 
2$\times$2$\times$3 supercells.

\subsection{Solution of the Boltzmann Transport Equation}

The second- and third-order IFCs obtained from the DFT calculations 
were used to solve the linearized Boltzmann Transport Equation (BTE): 
\begin{equation}
\bb{v}_\lambda \cdot \bb{\nabla}T \frac{\partial n^0_\lambda}{\partial T} =
\frac{dn_\lambda}{dt} \bigg{|} _{\mathrm{scatt.}}, 
\end{equation}
where $n^0_\lambda$ and $n_\lambda$ are the phonon distribution at 
and out of equilibrium, respectively, $\bb{v}_\lambda$ is the group velocity, and 
$dn_\lambda/dt \big{|} _{\mathrm{scatt.}}$ is the rate of change in the
phonon distribution as a result of phonon-phonon scattering.
We can assume the difference between $n_\lambda$ and $n^0_\lambda$ to be of the 
form
\begin{equation}
n_\lambda=n^0_\lambda-\bb{F}_\lambda\cdot \bb{\nabla}T \partial n^0_\lambda / \partial T.
\end{equation}
 Then, the linearized BTE can be rewritten in the following way~\cite{shengbte}
\begin{equation}
\bb{F}_\lambda=\tau^0_\lambda (\bb{v}_\lambda+\bb{\Delta}_\lambda) ,
\label{shengBTEform}
\end{equation}
where $\bb{F}_\lambda$ is the generalized mean free path,
 $\tau^0_\lambda$ is the relaxation time of mode $\lambda$
in the relaxation time approximation (RTA), 
and $\bb{\Delta}_\lambda$ gives the deviation of the solution
from the RTA. 
By taking into account only isotopic and anharmonic phonon scattering processes,
we can write $\tau^0_\lambda$ as
\begin{equation}
\frac{1}{\tau^0_\lambda}=\Gamma_\lambda=\Gamma_\lambda^{\mathrm{isot}}+\Gamma_\lambda^{\mathrm{anh}} ,
\end{equation}
where the total scattering rate of mode $\lambda$, $\Gamma_\lambda$, is the sum of the 
isotopic scattering rate $\Gamma_\lambda^{\mathrm{isot}}$ and the
anharmonic scattering rate $\Gamma_\lambda^{\mathrm{anh}}$.
Scattering caused by isotopic disorder depends on the mass variance 
of the elements in the compound and was included through the model due 
to Tamura~\cite{tamura}, for which we employed the natural abundances 
of Si and O (see Table~\ref{tabisot})~\cite{weast1988crc}.
Three--phonon scattering processes were included to account for the
anharmonic scattering, which is computed directly from the 
third-order IFCs obtained from the DFT calculations.
Higher order scattering processes were neglected, since they are 
expected to be important only at very high temperatures~\cite{pomeranchuk}.
\begin{table}
\begin{tabular}{cc|cc}
Isot.   & abund. (\%) &Isot. & abund. (\%)  \\ \hline
$^{28}$Si & 92.23  & $^{16}$O & 99.76 \\ 
$^{29}$Si &  4.67  & $^{17}$O &  0.038 \\ 
$^{30}$Si &  3.10  & $^{18}$O &  0.2 
\end{tabular}
\caption{Stable isotopes of Si and O and their relative naturally occurring abundances.}
\label{tabisot}
\end{table}

The BTE in the form given in Eq.~(\ref{shengBTEform}) is solved
iteratively~\cite{omini1995iterative} in a parameter--free approach
as implemented in the \textsc{ShengBTE} code~\cite{shengbte} to
obtain the lattice thermal conductivity tensor
\begin{equation}
\kappa_\ell^{\alpha\beta} = \frac{1}{k_B T^2 \Omega N} \sum_\lambda n^0 (n^0 + 1 )
(\hbar \omega_\lambda)^2 v_{\alpha,\lambda}F_{\beta,\lambda} ,
\label{eq:kappa}
\end{equation}
where $\alpha$ and $\beta$ are the three coordinate directions
$x$, $y$, and $z$; and $k_B$, $T$, $\Omega$ and $N$ are the Boltzmann
constant, the temperature, the volume of the unit cell and the number
of {\bf q}-points in the integral over the BZ, respectively. The sum 
runs over all the phonon
modes $\lambda$, $\hbar$ is the reduced Planck constant, and
$\omega_\lambda$ is the phonon frequency.

For the first iteration step $\bb{\Delta}_\lambda$ is set to zero, 
which is tantamount to starting the iterative procedure from the 
RTA solution. The solution is considered to be converged when the 
relative change in all the components of the thermal conductivity 
tensor becomes smaller than $10^{-5}$.

\subsection{Convergence issues}

The convergence tests of the relevant parameters of the 
electronic structure calculations --plane-wave energy cutoff,
{\bf k}-point mesh-- were carried out as usual. In the same
way, we carefully checked the dependence of the IFCs on the
supercell size, finding that a 3$\times$3$\times$4 supercell
yielded converged values.

Some convergence issues related to the numerical solution of 
the BTE must be addressed as well. The first one concerns
the $\delta$ functions appearing in the energy conservation terms in
the scattering rates, which are approximated by Gaussian functions for which
the standard deviation is computed in a self-adaptive manner as
implemented in the \textsc{ShengBTE} code~\cite{shengbte}.
The locally adapted Gaussian standard deviation appearing in the
computation of the scattering rates (both anharmonic and isotopic)
was scaled by 0.1. 
This scaling was necessary because unscaled standard deviations 
prevented to converge the solution of the BTE in a 
non-negligible subset of the targeted pressure-temperature values.
However, we verified in a few cases where the unscaled
calculation converged, that the employed 0.1 scaling yielded
minor quantitative differences in the thermal conductivity.

We also observed that solving the BTE in a too dense $\bb{q}$-point mesh 
could lead to an imaginary acoustic mode close to the critical pressure,
because of inaccuracy of the Fourier interpolation in the calculation of the phonon bands.
In order to avoid spurious effects due to such contributions we employed a slightly 
sparse {\bf q}-point mesh (6$\times$6$\times$6) so that no imaginary 
frequencies are sampled close to the critical pressure around the 
$\Gamma$ point in the BZ integrations. Although the exact values of 
the conductivity are slightly underconverged with the {\bf q}-point grid and 
hence not quantitatively accurate, particularly close to the lattice 
thermal conductivity peak at each temperature, we have checked that 
our main qualitative results remain valid with denser grids.

Finally, we stress that all the results discussed in this work have 
been obtained by solving the BTE iteratively. Nevertheless, we have 
found that the RTA results for the lattice thermal conductivity 
differ at most by 10\% in the whole $(T,P)$ region we explored.

\begin{figure}
 \includegraphics[scale=0.05]{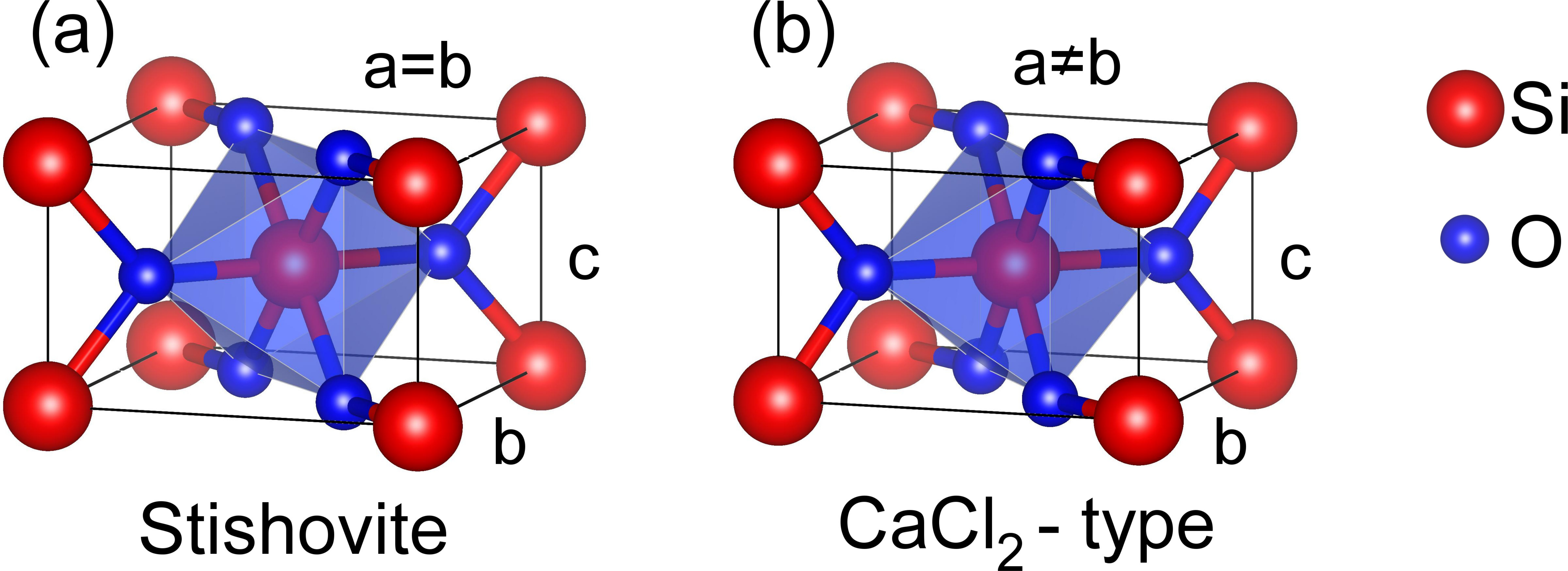}
 \caption{Unit cell of SiO$_2$  in the stishovite phase (a) and the CaCl$_2$-type phase (b). Si (O) atoms are 
shown in red (blue). The stishovite phase is a tetragonal crystal while the CaCl$_2$-type phase is
 orthorhombic. Note that the oxygen octahedron (highlighted in blue) is tilted in the CaCl$_2$-type phase.}
 \label{crystal}
\end{figure}

\section{Results}
\subsection{Phase transition under pressure}

\begin{figure}
 \includegraphics[scale=0.7 ]{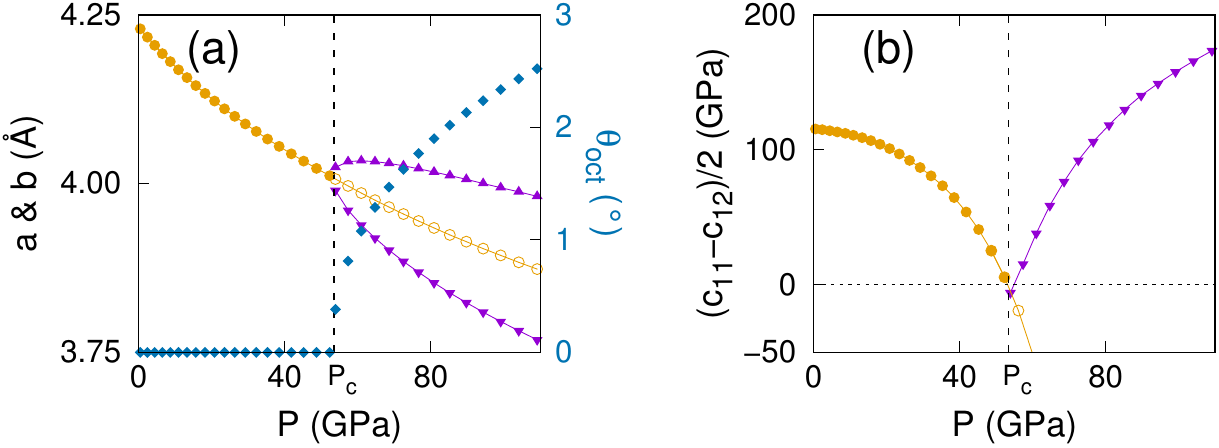}
 \caption{(a) Lattice parameters $a$ and $b$ for SiO$_2$  as a function of pressure as obtained with DFT.
 Results for the stishovite phase are shown in orange (circles) while those for the CaCl$_2$-type 
phase are shown in purple (triangles). The rotation angle of the oxygen octahedron is shown with unconnected blue points.
(b) Dressed shear modulus $\frac{(c_{11}-c_{12})}{2}$ as a function of pressure for the stishovite
 (orange) and CaCl$_2$-type (purple) phases. The shear modulus vanishes close to the phase transition.}
 \label{angles}
\end{figure}

The crystal structure of the two silica polymorphs studied is shown in Figure~\ref{crystal}. The stishovite
 phase belongs to the tetragonal crystal family with space group $P4_2/mnm$, while the CaCl$_2$-type phase belongs
 to the orthorhombic crystal family with space group $Pnnm$. The latter differs from the former in a tilt of 
 the oxygen octahedra surrounding the Si atoms and has lattice parameter $b\neq a $.
 We computed the relaxed lattice parameters and atomic positions of silica for pressures ranging from 0 to 100 GPa
 in both phases. In the stishovite phase we imposed $b=a$, while this restriction was lifted for the CaCl$_2$-type phase.
 The calculated $a$ and $b$ lattice parameters are shown in Figure~\ref{angles}, along with the oxygen
 octahedra rotation angle. At the critical pressure,
 $a$ starts to differ from $b$, and the oxygen octahedra become tilted in the CaCl$_2$-type phase.

 This phase transition is known to be driven
 by the softening of the optical $B_{1g}$ phonon mode in the BZ center, which corresponds
 precisely to a pure rotation of the oxygen octahedra. The critical pressure $P_c$ is taken as that for which the
 dressed shear modulus $(c_{11}-c_{12})/2$ (where the $c_{ij}$ is the $ij$ component of the stiffness tensor
 in Voigt notation) vanishes in the stishovite phase, \textit{i.e.}, the pressure
 at which the system shows no resistance to the shearing of the cell and the oxygen octahedra rotation.
 The computed shear modulus as a function of
 pressure is shown in Figure~\ref{angles} (b). We obtain $P_c$=53.4 GPa, in good agreement 
 with previous calculations~\cite{togo2008first} and experiments~\cite{kingma1995transformation}. 
This second-order phase transition is of the pseudoproper ferroelastic 
type, since the driving order parameter is not strain itself, but the 
strain emerging at the phase transition has the same symmetry as that 
order parameter~\cite{ferroic,Carpenter,togo2008first}.

\begin{figure}
 \includegraphics[scale=0.82]{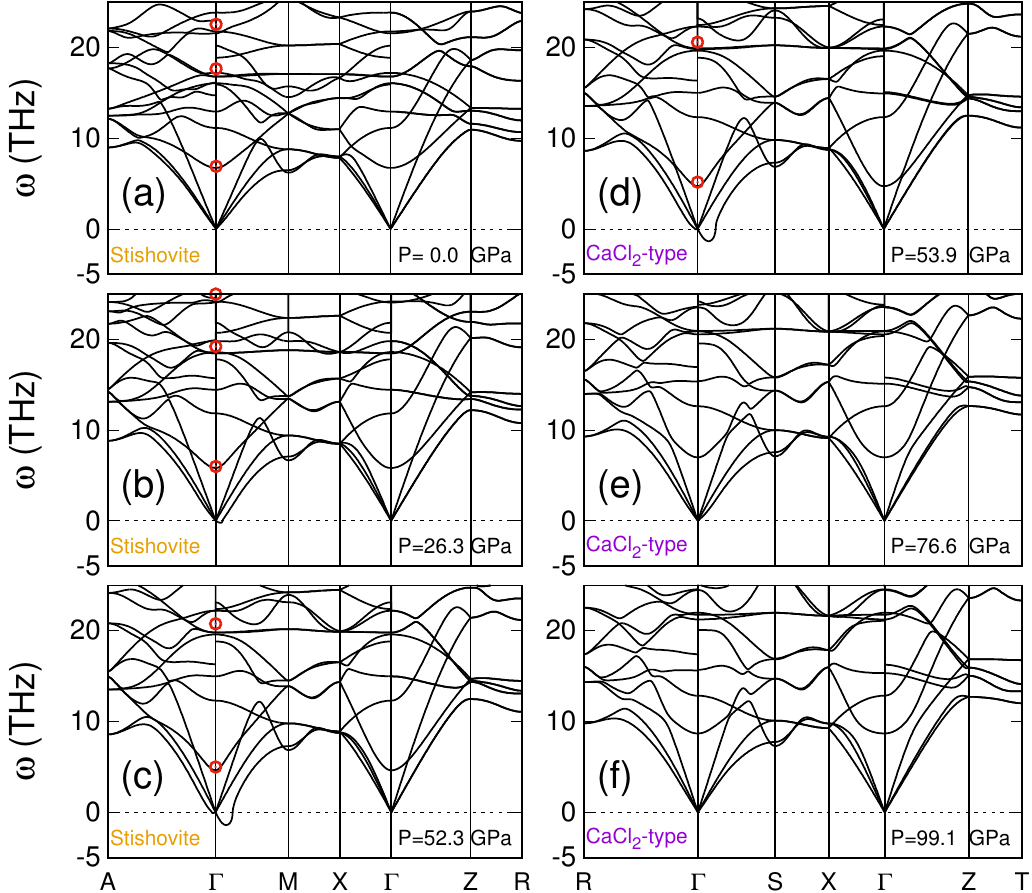}
 \caption{Phonon band structure for three representative pressures of the stishovite phase --(a) to (c)--
 and the CaCl$_2$-type phase --(d) to (f)--. The modes have been computed for $P$=0, 26.3, 52.3, 53.9, 76.6 and 99.1 GPa.
 Red circles correspond to experimental values for Raman-active modes (only available for pressures up to 60 GPa) taken from Ref[\onlinecite{heinz1995new}].
} 
 \label{phonon}
\end{figure}

Next, we compute the phonon band structure of silica in the stishovite and CaCl$_2$-type phases at different pressures.
 Some representative cases of both phases are shown in Figure~\ref{phonon}. We first note that some non--degenerate
 modes (such as the 3$^\text{rd}$ and 4$^\text{th}$ optical modes) of the CaCl$_2$-type phase become degenerate in the 
stishovite phase due to the higher symmetry of silica in the tetragonal phase. Moreover, the phonon bands generally gain energy
 with increasing pressure, the only two exceptions being the lowest optical band around $\Gamma$ (associated to the $B_{1g}$ mode
responsible for the structural phase transition), which softens close to $P=P_c$ and 
 then gains energy together with the rest of the modes, and the
 lowest energy band in the high symmetry point $A$ ($S$) in the stishovite (CaCl$_2$-type) phase 
(see video in Supplemental Material~\footnote{See Supplemental Material at [URL] for a video of the phonon band structure evolution of SiO$_2$ with pressure}).
 For pressures between 25 and 70 GPa, one of the acoustic bands becomes imaginary in a small region of the BZ       
 along the  $\Gamma-M$ direction for the stishovite phase~\cite{togo2008first,tsuchiya}, and along
 the $\Gamma-S$ direction for the CaCl$_2$-type phase.
 This artifact is sometimes attributed in the literature to the finite supercell size employed in the calculations
 (see for instance Ref.~[\onlinecite{togo2008first}]), but our tests for SiO$_2$ and other materials
 (\textit{e.g.} BaTiO$_3$) seem to suggest that the problems persist even if considerably larger supercells are used.
 Instead, the appearance of imaginary acoustic modes close to the Brillouin zone center seems to be related
 with a systematic error due to the so-called Fourier interpolation of the dynamical matrix for non-commensurate
 {\bf q}-points. Naturally, this issue becomes more obvious (even pathological) in regions where we have phonons with
near-zero frequencies (as \textit{e.g.} in the center of the Brillouin zone), since in such cases a problem in the
 interpolation scheme will typically result in some frequencies becoming imaginary.
 Thus, the calculation of the integrals in reciprocal space appearing in the BTE must be done with care, avoiding 
 the small region where this acoustic band becomes imaginary 
 so that no unphysical results are obtained.

 \subsection{Thermal conductivity}
 \subsubsection*{Solution of the BTE in the $(T,P)$ space}

\begin{figure}
 \includegraphics[scale=0.8]{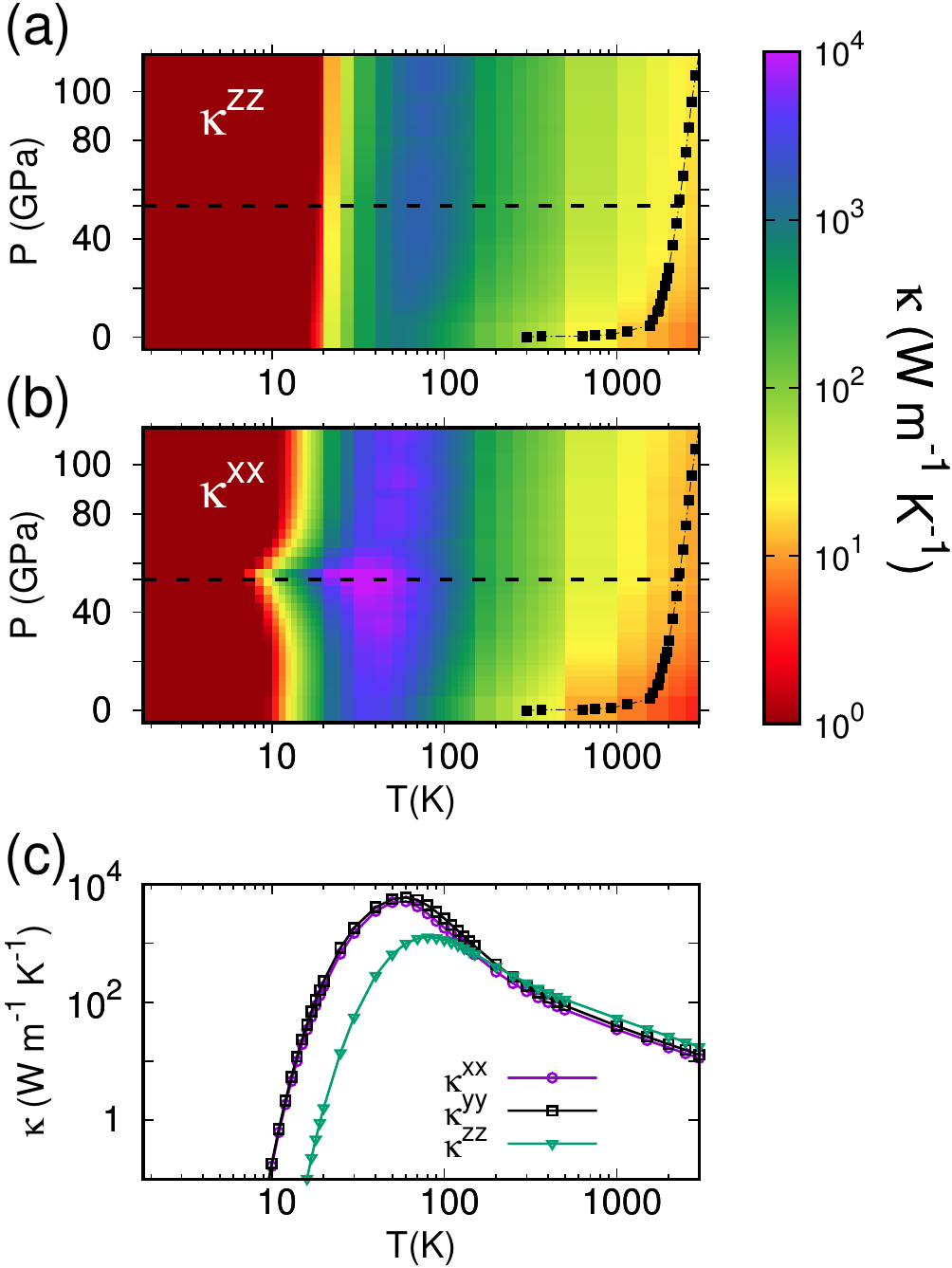}
 \caption{Lattice thermal conductivity of SiO$_2$  for different pressures and temperatures (see color code on the right).
 The $zz$ component of the lattice thermal conductivity is shown in (a), and the $xx$ component in (b).
 The results for the stishovite phase are shown only for pressures below $P_c$, while the results for 
 the CaCl$_2$-type phase are displayed above $P_c$. The $zz$ component shows a slight increase as pressure increases,
 but no special feature is seen at $P=P_c$. The $xx$ component, is larger than the $zz$ one for $T<$200~K,
 showing a huge enhancement around the critical pressure for temperatures below 40~K.
 Dotted dashed black curves show the average geotherm according to Dziewonski and Anderson's Earth model~\cite{geothermal}.
 (c) Lattice thermal conductivity along the P=76.6~GPa isobar. Purple circles, black squares, and green triangles
 correspond to the $xx$, $yy$, and $zz$ components, respectively.
}
 \label{kappatop}
\end{figure}

We next solve the BTE for both silica phases for several values of temperature, $T$, and pressure, $P$.
 Figures~\ref{kappatop} (a) and (b) show the lattice thermal conductivity components $\kappa^{zz}$ and 
$\kappa^{xx}$, respectively.
  We find that the off--diagonal components are negligible in the whole $(T,P)$ space explored 
 and that the $\kappa^{yy}$ component
 (not shown) is equal to the $\kappa^{xx}$ component in the stishovite phase due to symmetry.
 In the CaCl$_2$-type phase $\kappa^{yy}$ shows a similar behavior to $\kappa^{xx}$.
 For pressures
 below (above) the critical pressure we only show the results for the stishovite (CaCl$_2$-type phase), \textit{i.e.},
 only the results for the stable phase at each pressure are shown.
 We first note that both components are substantially different in most of the $(T,P)$ space. 
More importantly, we find an 
 astonishingly large peak in $\kappa^{xx}$ close to the critical pressure at both sides
 of the critical line for temperatures between 8 and 70~K.

In Figure~\ref{kappatop} (c)
 we show the lattice thermal conductivity for the P=76.6~GPa isobar. We find the characteristic trend
 of the lattice thermal conductivity curves for bulk systems. At low temperatures $\kappa$ increases
 with $T$ until it reaches a maximum, after which the conductivity decreases more slowly. We also note that 
 $\kappa^{xx}$ is almost equal to $\kappa^{yy}$ in the whole temperature range, while $\kappa^{zz}$ is smaller
 below $T=200$~K. Thus, the thermal response of silica is clearly anisotropic in general.

\begin{figure}
 \includegraphics[scale=1.5]{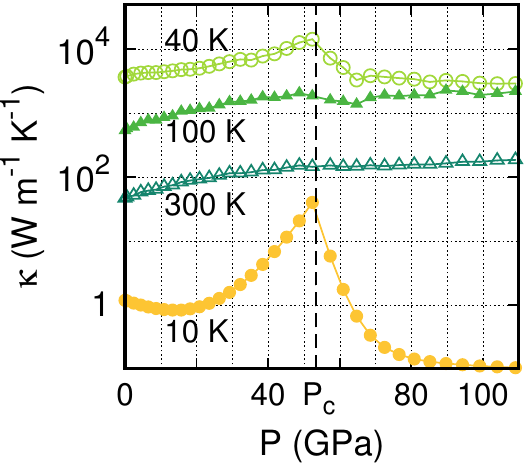}
 \caption{Constant temperature profiles of $\kappa^{xx}$ as a function of pressure, 
for $T$=10~K (filled circles), $T$=40~K (empty circles), $T$=100~K (filled triangles) and $T$=300~K (empty triangles).
}
 \label{kappap}
\end{figure}

 In Figure~\ref{kappap} we show 
 $\kappa^{xx}$ profiles for different isotherms. The lattice thermal conductivity close to $P=P_c$
 increases up to 2 orders of magnitude for low temperatures (see the $T$=10~K line in Fig.~\ref{kappap}),
 and about 1 order of magnitude for medium temperatures (see the 40~K isotherm in the same figure).
 At higher temperatures, at which 
 the scattering is dominated by anharmonic processes, the peak is smoothed
 out until it disappears (see the 100~K and 300~K isotherms).
This suggests that the peak stems from a shift in the dominant scattering
 processes, from isotopic to anharmonic phonon-phonon, as temperature increases
 (recall that we are calculating the bulk lattice thermal conductivity, thus ignoring boundary scattering
 processes).
This can be confirmed, for example, by
 recomputing the thermal conductivity maps $\kappa(T,P)$
 with no anharmonic scattering. Since isotopic scattering is dominant
 at low temperatures, we obtained almost identical results for $T<$100~K, while at each
 pressure $\kappa(T)$ saturates to a constant
 value for $T\gtrsim$150~K.

\subsubsection*{Origin of the peak around the critical pressure}
\begin{figure}
 \includegraphics[scale=0.70]{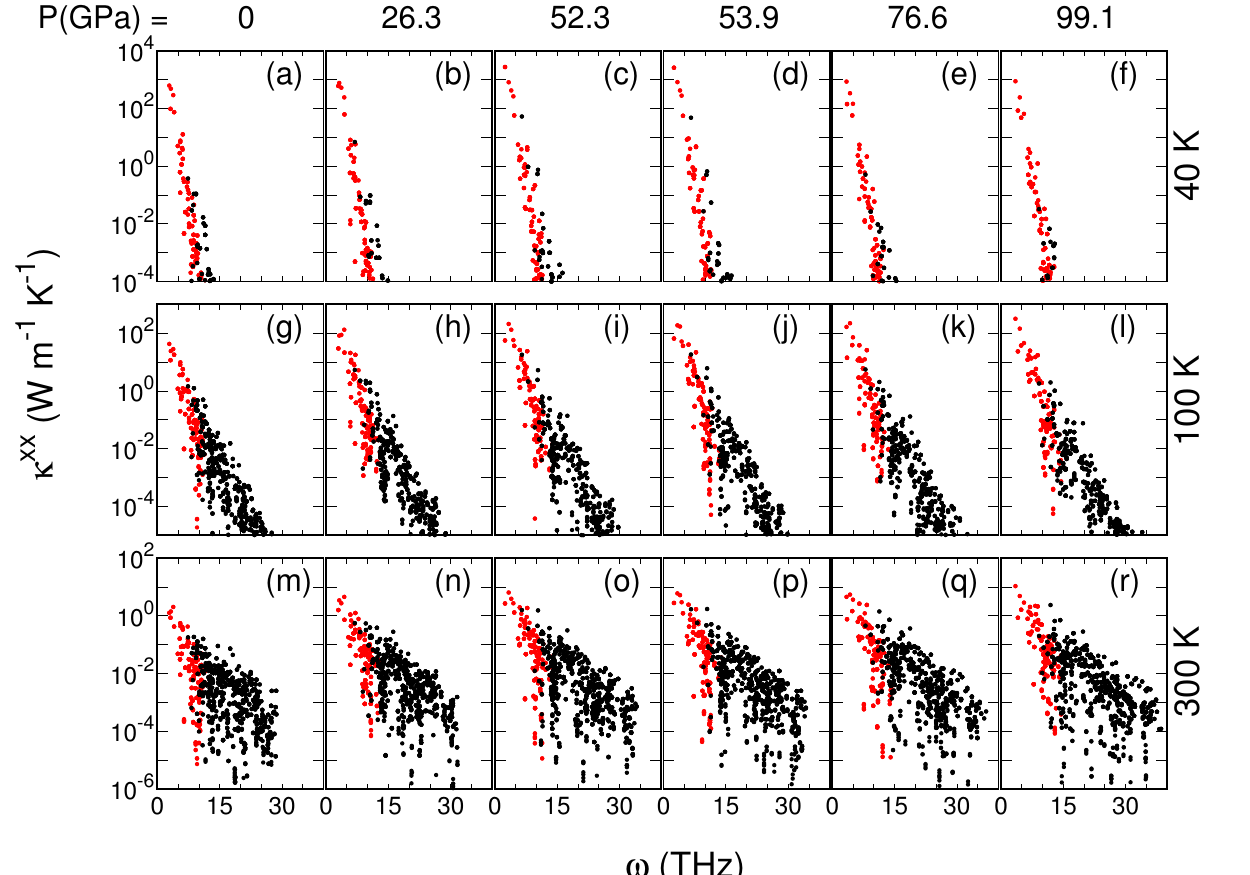}
 \caption{Contribution to the total lattice thermal conductivity for each mode $\lambda$ ($\kappa_\lambda$).
 The first, second and third rows correspond to calculations at $T$=40, 100 and 300~K respectively.
 The columns from left to right correspond to $P$=0, 26.3, 52.3, 53.9, 76.6 and 99.1 GPa, where the results are 
 shown only for the stable phase at each pressure. Close to the critical pressure, a few acoustic modes
 are responsible for the anomalously large value of $\kappa^{xx}$  at $T$=40~K --(c) and (d)--.
 For $T$=100~K --(g) to (l)-- and $T$=300~K --(m) to (r)--, there is no peak in
  $\kappa^{xx}$ at $P=P_c$ and the $\kappa_\lambda$ of the
 highest contributing modes increases 
monotonically with pressure. Red (black) dots correspond to acoustic (optical) modes. 
} 
 \label{bothkappaPT}
\end{figure}

 In order to trace the origin of the notorious peak in the in--plane thermal conductivity, we analyze
 the contribution to $\kappa^{xx}$ of each
 mode in each of the sampled points in the BZ.
 The individual contributions of all the modes included in our BZ integrations at different
  representative temperatures and pressures are depicted in Figure~\ref{bothkappaPT}. For each {\bf q}-point in the
 irreducible BZ the contribution of the 18 modes is shown. At $T$=40~K (upper row in the Figure),
 we see that $\kappa^{xx}$ is strongly dominated by the contribution of a few low energy modes. Also, one (degenerate)
 mode contributes with up to 2.7$\cdot 10^{3}$~W m$^{-1}$K$^{-1}$ for $P\approx P_c$ but is not so active
 at high or low pressures.
 For higher temperatures ($T$=100~K and $T$=300~K, second and third rows in the Figure respectively),
 although the dominant contributions still correspond to low energy modes, optical phonons start to play an increasingly
 important role in the total thermal conductivity. Furthermore, the general trend is a monotonic increase
 of the individual contributions with increasing pressure.
 In this way, we can attribute the peak in the in--plane lattice thermal
 conductivity at low temperatures to the enhancement of the contribution of particular low energy acoustic modes.

\begin{figure}
 \includegraphics[scale=0.95]{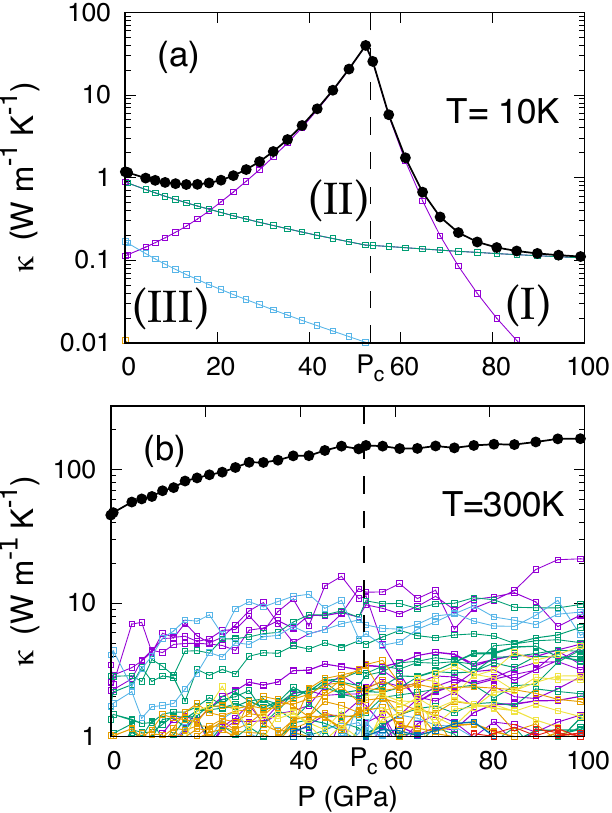}
 \caption{Contribution to $\kappa^{xx}$ as a function of pressure of the different modes for (a) $T$=10~K and (b) $T$=300~K.
  The total thermal conductivity at each temperature is shown in black circles.}
 \label{modespressure}
\end{figure}

We now analyze the evolution of the individual contributions $\kappa^{xx}_\lambda$ with pressure. The results for two 
representative temperatures (10 and 300~K) are shown in Figure~\ref{modespressure}. 
The total $\kappa^{xx}$ is also presented with black filled circles for comparison.
On the one hand,
 at $T$=10~K thermal transport is dominated by essentially one mode --purple squares in the Figure, mode~(I)--
 for pressures between 30 and 60~GPa (see that the total $\kappa^{xx}$ is almost equal to the
 contribution of this mode in this pressure range).
 The aforementioned mode is a low energy acoustic phonon
 with wavevector $\boldsymbol{q}=(1/6,1/6,0)$ in reciprocal vector units, which is representative of an entire 
region of low energy modes close to the BZ center.
 Away from this pressure window, the contribution of another acoustic
 mode --mode~(II)-- with $\boldsymbol{q}=(1/3,0,0)$ in reciprocal vector units (green squares) becomes larger than that 
of mode~(I), hence dominating the thermal conductivity at low and high pressures. The contribution of the 
next highest contributing mode is about one order of magnitude smaller, thus thermal transport in silica
 close to the phase transition is strongly governed by the behavior of the acoustic phonons represented by mode~(I).
On the other hand, at $T$=300~K many modes give contributions in the range of 1 to 10 W m$^{-1}$K$^{-1}$. The total 
 $\kappa^{xx}$ is thus the addition of many contributions of similar magnitude, and no particular mode dominates the
 thermal transport properties of SiO$_2$  in the studied pressure range. Moreover, there is only a minor dip in the
 total thermal conductivity close to the critical pressure, and this feature can not be attributed
 to an individual mode.
 Although the overall thermal conductivity increases almost monotonically with pressure,
 not all of the highly contributing modes follow this behavior. 

\begin{figure}
 \includegraphics[scale=0.7]{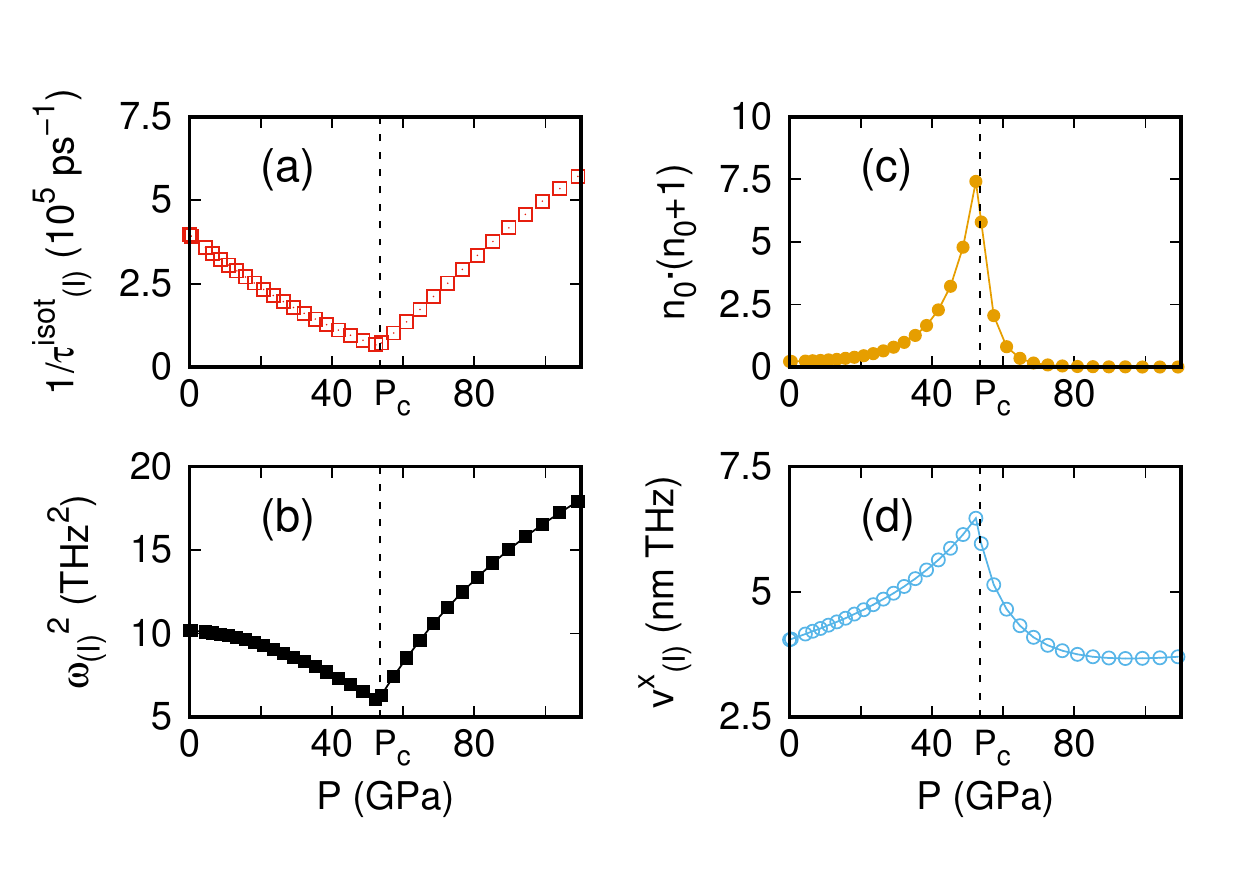}
 \caption{Evolution of isotopic scattering rate of mode~(I), squared frequency of mode~(I), $n_0(n_0+1)$ at $T=$10~K for mode~(I) and
$x$ component of the group velocity of mode~(I)
 with pressure --(a), (b), (c) and (d) respectively--. The critical pressure $P_c$ is indicated with a vertical dashed line 
as a guide tot he eye in all cases.
}
 \label{follow}
\end{figure}

 With the aim of acquiring some deeper insight of the origin of the peak, we now analyze
the different factors that determine the contribution of mode (I) and its evolution
with pressure.
 Figure~\ref{follow} shows the evolution with pressure of the main different factors
 involved in $\kappa_{(I)}^{xx}$, namely the isotopic scattering rate, the square of the frequency,
 the product of populations $n^0_{(I)}(n^0_{(I)}+1)$ and the group velocity --see Eq.~(\ref{eq:kappa})--. 
 While the term $\omega_{(I)}^2$ decreases by a factor
 of 0.6 at the critical pressure (with respect to $P=0$), the remaining pressure dependent factors increase as the pressure
 approaches $P_c$. In fact, a decreasing $\omega_\lambda^2$ is always overcompensated by the growth of $n^0_\lambda (n^0_\lambda + 1)$,
\textit{i.e.}, the softening of a phonon band results in a larger contribution to the thermal conductivity 
 (not taking into account the possible changes in the relaxation times). The largest ratio for mode~(I) is precisely
 that of the phonon population product  $n^0_\lambda (n^0_\lambda + 1)$, which accounts for a factor of 33.5 in the enhancement 
 close to the critical pressure. The isotopic scattering rate (which at low temperatures is much larger
 than the anharmonic scattering rate) decreases by a factor of 6, increasing
 the contribution of mode~(I) to the thermal conductivity by the same factor and being this the second largest ratio.
 The decrease in the unit cell volume and the increase in group velocity close to $P_c$ contribute to the total ratio 
with factors of 1.1 and 2.6 respectively. The total enhancement factor is of 338.4. 

\subsubsection*{Thermal conductivity of SiO$_2$  along the geotherm}
\begin{figure}
 \includegraphics[scale=0.70]{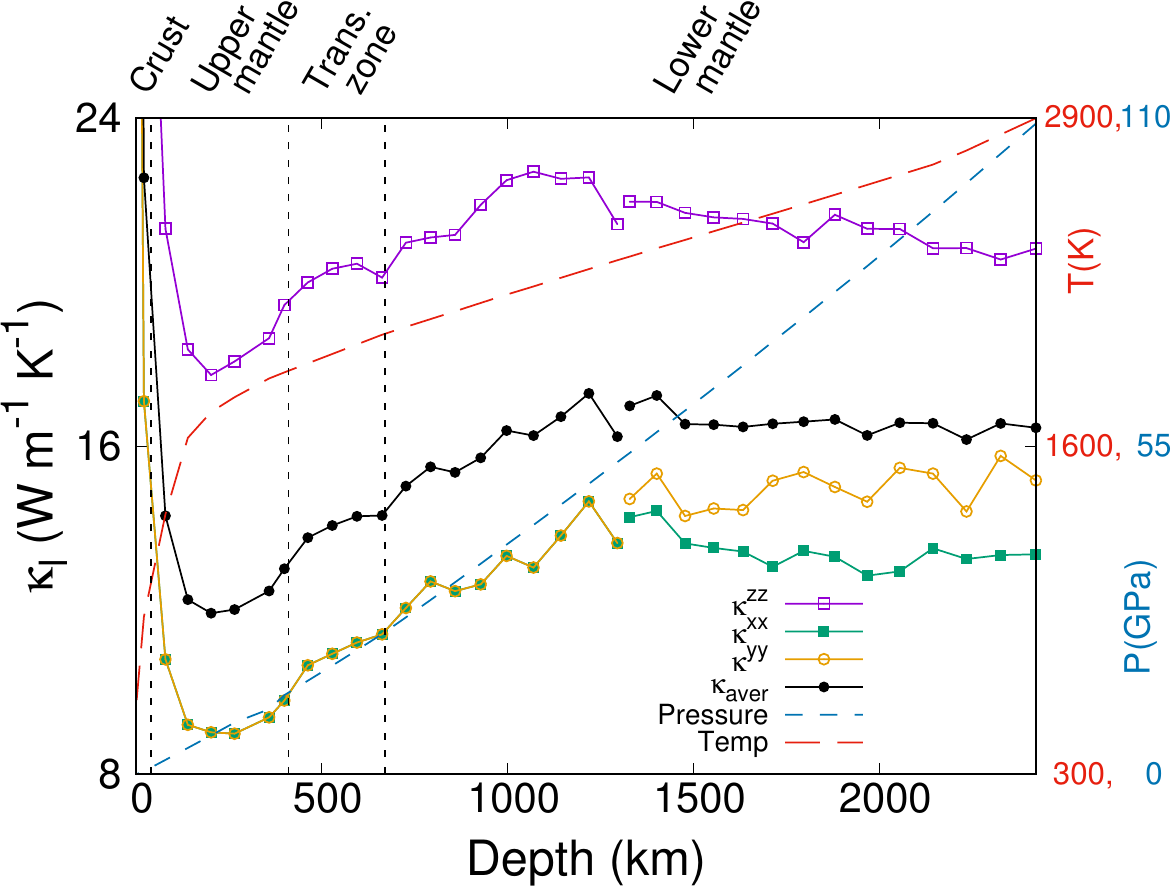}
 \caption{Lattice thermal conductivity of silica in the stishovite and CaCl$_2$-type phases along an average geotherm.
 The results for the stishovite (CaCl$_2$-type) phase are shown for $P<(>) P_c$, and the phase transition is marked by a discontinuity in the lattice thermal
 conductivity lines. Filled squares, empty circles and empty squares correspond to 
 the diagonal components $\kappa^{xx}$, $\kappa^{yy}$ and $\kappa^{zz}$ respectively, and the average lattice thermal conductivity
 $\kappa_{\mathrm{aver}}=\frac{1}{3}(\kappa^{xx}+\kappa^{yy}+\kappa^{zz})$ is shown with filled circles (see legend).
 The average estimated temperature (pressure) at each depth from the Earth's crust is shown with a long--(short--)dashed red (blue) line.
 There is a marked decrease in the three diagonal components of $\kappa$ close to $P=P_c$.
} 
 \label{geotherm}
\end{figure}

The thermal conductivity of the Earth's mantle determines its convection 
mechanism and the heat budget of our planet~\cite{MgO2}. 
Owing to the great interest of SiO$_2$ in the field of geophysics, 
we additionally computed the thermal conductivity tensor of silica along an 
average geotherm~\cite{geothermal} from the Earth's crust up to 
depths of 2400~km, deep inside the lower mantle. The results are 
displayed in Figure~\ref{geotherm}. We first note that for high 
temperatures the anisotropy is reversed, being $\kappa^{zz}$ larger 
than $\kappa^{xx}$ and $\kappa^{yy}$, as opposed to the results 
analyzed above for lower temperatures. In the stishovite phase, 
the three diagonal components increase approximately linearly with 
depth up to the critical pressure, while for the CaCl$_2$-type phase 
the thermal conductivity is roughly constant with depth.
Nevertheless, a decrease in all the components is seen close
to the phase transition, which is marked by a discontinuity in the 
lines in the Figure. Interestingly, we note that in works by Murphy~\textit{et. al.} 
on PbTe-based materials~\cite{Ronan,Ronan2} a dip in the thermal conductivity 
of ferroelectric compounds was also predicted close to a structural 
phase transition.

\subsection{Discussion}

The present work was partly motivated by our conjecture that SiO$_{2}$
should be a suitable model compound to investigate changes in thermal
conductivity across a soft-mode-driven structural phase
transition. Indeed, it seemed to us that SiO$_{2}$'s
 transformation between the stishovite and CaCl$_{2}$-type structures
 might be representative of
other soft-mode ferroelastic (and ferroelectric) transitions, with the
advantage that the transition-controlling parameter (pressure, as
opposed to temperature) is trivial to handle in a first-principles
simulation. Hence, let us briefly discuss the generality of our
findings for SiO$_{2}$, and whether our initial conjecture was
correct.

Most importantly, our simulations predict that SiO$_{2}$ will display
some strong changes in its low-temperature thermal conductivity around
the transition pressure. Interestingly, such features are not directly
related with the optical soft mode itself, which is the primary order
parameter for the transformation but has a negligible impact in
$\kappa$. Instead, they can be traced back to the accompanying
softening of an acoustic band, an effect that is specific to the
peculiar pseudoproper ferroelastic character of the
investigated transition in SiO$_{2}$. Hence, this suggests
that enhancements of the low-temperature conductivity, as those we
observed in SiO$_{2}$, can be expected to occur at ferroelastic
transitions involving a significant softening of the acoustic
bands. Our results also suggest that, in contrast, other soft mode
transformations -- as, \textit{e.g.}, ferroelectric or anti-ferrodistortive
transitions in which only an optical bands softens -- will in general
not present such marked effects in the thermal conductivity.

Additionally, our results show that, in SiO$_{2}$, significant changes
in the lattice thermal conductivity are restricted to relatively low
temperatures. More precisely, we find that, while the softening of the
key acoustic band causes a large increase in its population at low
temperatures --Fig.~\ref{follow} (c)--, the relative population change is small at
higher values of $T$; as a result, the effect in the conductivity at
higher temperatures is modest as well. Naturally, we do not know
whether, in other materials, phonon frequencies and populations will
control the variations of $\kappa$ to the same extent that we find
they do in SiO$_{2}$. Yet, it is clear that, unlike the
material-specific scattering rates or group velocities in Eq.~(\ref{eq:kappa}),
temperature-dependent populations are governed by universal
laws. Therefore, the softening of a low frequency mode (either optical
 or acoustic) will in general enhance the low temperature thermal conductivity.

Nevertheless, let us note that, at higher temperatures, other factors can play a
more dominant role. For example, as shown in Refs.~\onlinecite{Ronan} and ~\onlinecite{Ronan2},
when three-phonon scattering becomes dominant, a mode softening can open (close)
phonon-phonon scattering channels, leading to significant reductions (enhancements)
of the thermal conductivity.

\section{Conclusions}
By means of state-of-the-art calculations, we have characterized the
stishovite and CaCl$_2$-type high pressure phases of silica and computed a
thermal conductivity map $\kappa(T,P)$ for a large number of points in
the $(T,P)$ space. We found that, for pressures close to the phase
transition, a large peak in the in-plane conductivity appears at both
sides of the phase transition and at $T$ below 70~K. Moreover, this
peak is not present in the out-of-plane conductivity, thus leading to
a highly anisotropic thermal material. We have tracked down the origin
of this unexpected peak, and found that it originates in the softening
of an acoustic band that becomes highly populated close to the phase
transition. We have discussed the general implications of our results,
which suggest that lattice thermal conductivity effects
associated to soft-mode transitions will be restricted to low
temperatures and, thus, may not be promising for applications at
ambient conditions. Finally, we have computed the thermal conductivity
along an average geotherm for its possible applications in the field
of geophysics, since SiO$_2$  is the most abundant compound in the
Earth's mantle.

\section*{Acknowledgments}
We acknowledge financial support by the Ministerio de Econom\'ia,
Industria y Competitividad (MINECO) under grant FEDER-MAT2013-40581-P
and the Severo Ochoa Centres of Excellence Program under grant
SEV-2015-0496, and by the Generalitat de Catalunya under grant
no. 2014 SGR 301. We also acknowledge the financial support of the
Luxembourg National Research Fund (Grant FNR/P12/4853155/Kreisel
COFERMAT). We acknowledge the use of computational resources of CESGA,
and the i2BASQUE academic network. Finally, we acknowledge Prof. Robert
T. Downs for pointing to Reference~[\onlinecite{geothermal}].

\end{document}